\documentclass[12pt,twoside]{article}
\usepackage{fleqn, espcrc1}
\usepackage{epsfig}

\usepackage{amssymb}

\title{HBT and Fluctuations: Recent Results }

\author{Sergey Y. Panitkin\address{Kent State University, Kent OH,
44242 USA}\thanks{present address: Brookhaven National Lab, Upton NY,
11973 USA}}

\begin{document}

\maketitle

\begin{abstract}
% Text of abstract
We give an overview of the latest results of HBT and fluctuations
studies in heavy ion collisions presented during the Quark Matter 2001 Conference.
\end{abstract}

%\begin{keyword}

%\PACS 
%\end{keyword}
%\end{frontmatter}

% main text
\section{Introduction}
\label{intro}
The study of small relative momentum correlations, a technique also
known as HBT~\cite{hbt} interferometry, is one of the most powerful
tools at our disposal to study complicated space-time dynamics of
heavy ion collisions~\cite{reviews}.  
It provides crucial information which helps to improve our
understanding of the reaction mechanisms and to constrain theoretical
models of the heavy ion collisions. It is also considered to be a
promising signature~\cite{rischke_96,lastcall} of the Quark Gluon Plasma
(QGP). Interpretation of the extracted HBT parameters in
terms of source sizes and lifetime is more or less straightforward for
the case of chaotic static sources. In the case of expanding sources
with strong space-momentum correlations (due to flow, etc.)
the situation is more difficult, but the concept of
length of homogeneity~\cite{makhlin_sinyukov} provides a useful
framework for the interpretation of data.\\
\section{HBT Results}
\label{hbt}
%CERN Results
New data from NA45~\cite{appelshauser} (CERES Collaboration) and
NA49~\cite{blume} experiments at CERN shed new light on the evolution
of the reaction dynamics at SPS energies. 
Both collaborations showed results of the systematic studies of the
HBT parameters in Pratt-Bertsch parameterization~\cite{pratt,bertsch}
for Pb+Pb(Au) collisions at 40A GeV.
Taking into account preliminary status of both analyses the agreement
between them is fairly good. 
The NA49 Collaboration analysis showed a striking similarity (including detailed
$K_T$ dependence, where $K_T$ is the average transverse momentum of a pair) 
between the radius parameters at 40 and 160A GeV. This, coupled with
results of the transverse radial flow analysis~\cite{nuxu}, suggests
that the reaction dynamics, despite the difference in the initial
energy density and final state multiplicities, are rather similar.\\
%----------------------------------------------------------------------
%-------------------------------------------------------------------
 \begin{figure}
\begin{center}
 \epsfig{file=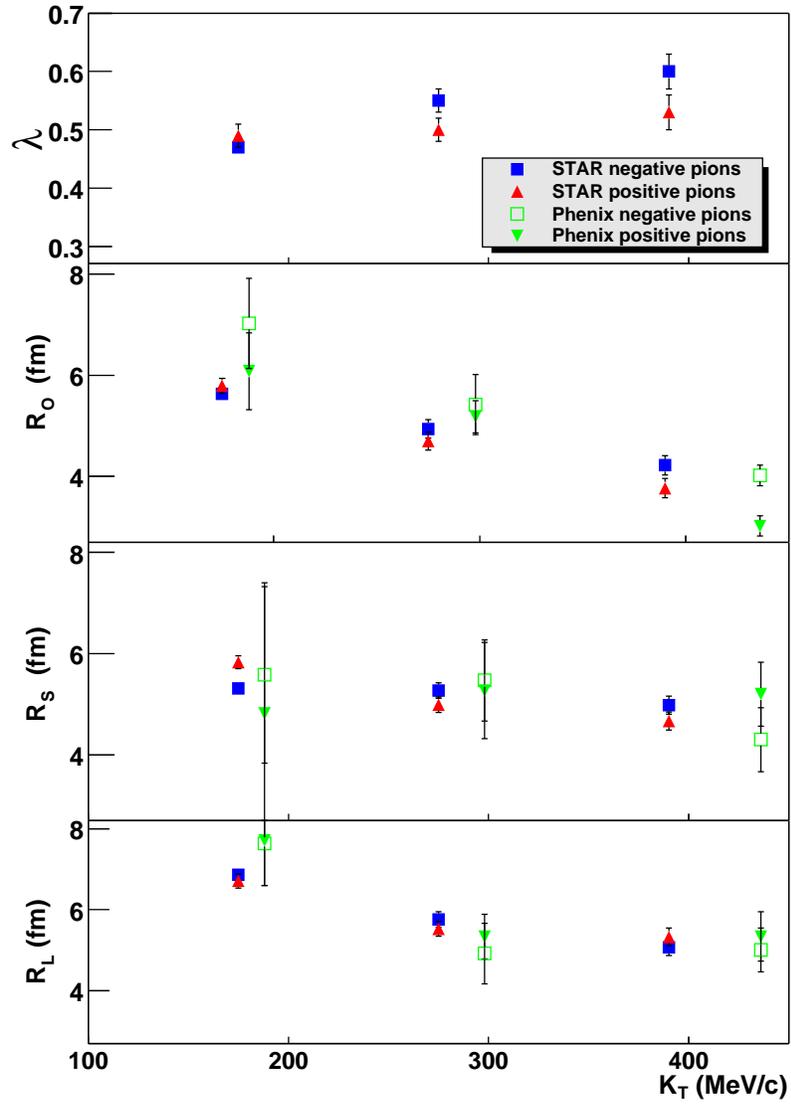,width=12.0cm}
 \caption{Preliminary data on $K_{T}$ dependence of pion HBT
 parameters reported by STAR~\cite{laue} and PHENIX~\cite{johnson}
 Collaborations.}   
 \label{panitkin_2}
\end{center}
 \end{figure}   
The STAR Collaboration has shown~\cite{laue} preliminary results
of the extensive analysis of the pion (both $\pi^+$ and $\pi^-$)
correlation functions measured at RHIC in Au+Au collisions at
$\sqrt{S_{NN}}$=130 GeV.  While an increase in radius parameters with
charged particle multiplicity was more or less expected, the observed
$K_T$ dependence, shown in Fig.~\ref{panitkin_2}, is somewhat of a
surprise. The general trend of the HBT radius parameters with $K_T$ is
consistent with strong space-momentum correlations due to transverse
flow, however, the $K_T$ 
dependence of the ratio of $R_O$/$R_S$ contradicts the model
predictions~\cite{rischke_96,soff_bass_dumitru}. Model calculations
predict the ratio to be greater than unity due to system lifetime
effects which cause $R_O$ to be larger than $R_S$. They also predict
that the $R_O$/$R_S$ ratio increases  with $K_T$. Such an increase seems to
be a generic feature of the models based on the
Bjorken-type~\cite{bjorken}, boost-invariant expansion scenario.  
Hence, it was surprising to see that the experimentally observed
ratio is less than unity and is decreasing as a function of
$K_T$. Currently, it is far from clear what kind of scenario can lead
to such a puzzling $K_T$ dependence. In principle, the observed behaviour
is consistent with the creation of an opaque
source~\cite{heiselberg_visher} in the collisions. However, to address this 
particular possibility seriously one needs to perform a detailed
analysis of the pion correlation function in Yano-Koonin-Podgoretski
parameterization~\cite{ykp}.\\

Preliminary results of the HBT analysis presented~\cite{johnson} by
the PHENIX Collaboration agree with the STAR results within
error bars but do not provide a sufficient cross check due to the
limited statistics of the first analysis. Clearly, further cross checks
of these results by other RHIC experiments are very desirable.\\
%-------------------------------------------------------------------
 \begin{figure}
\begin{center}
 \epsfig{file=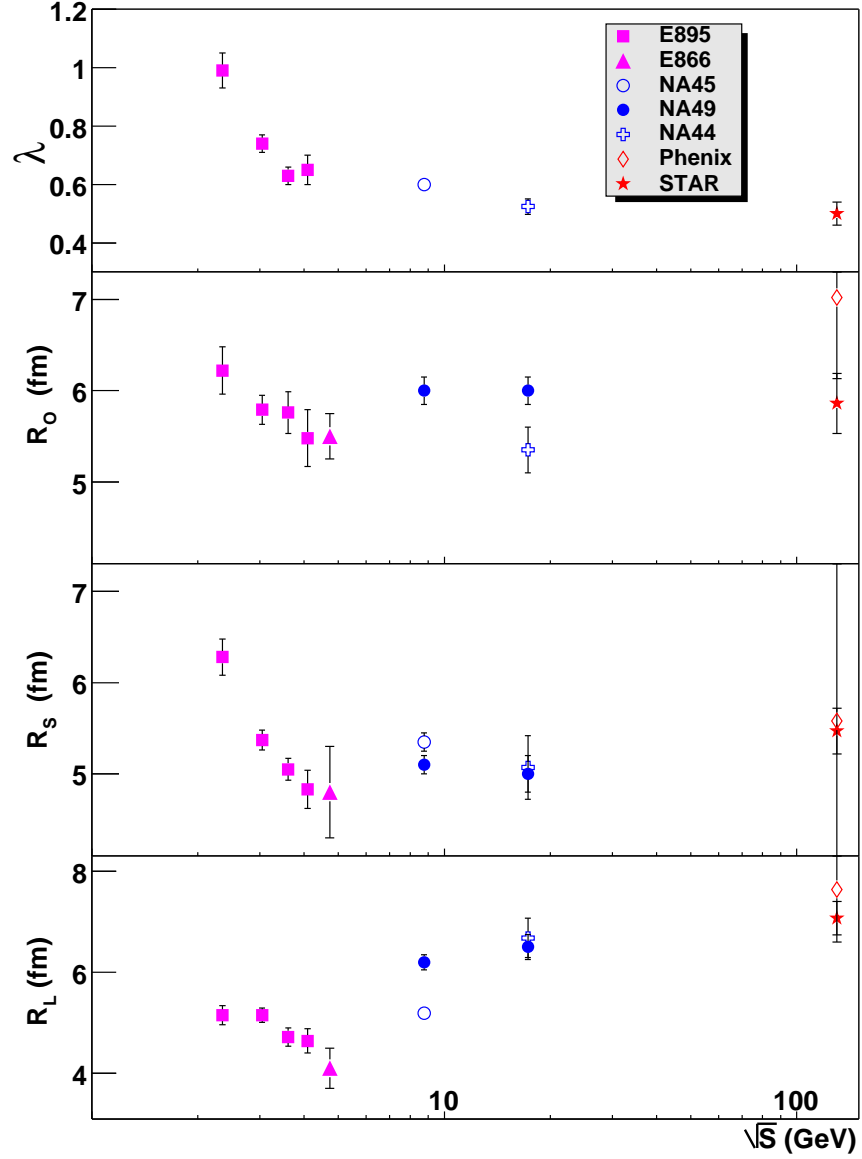,width=12.0cm}
 \caption{Measured Pratt-Bertsch parameters of
 two-particle correlation functions from central heavy-ion collisions at
 different
 energies~\cite{appelshauser,blume,laue,johnson,e895,e866,na44}. See
 text for detailes.}  
 \label{panitkin_1}
\end{center}
 \end{figure}
New results from RHIC and lower SPS energy significantly extended HBT
systematics. 
Figure~\ref{panitkin_1} shows evolution with energy of the
Pratt-Bertsch parameters of the two-particle correlation functions
measured~\cite{appelshauser,blume,laue,johnson,e895,e866,na44} in central heavy-ion
collisions. It is evident from  the figure that HBT parameters change
most rapidly at the AGS energies ($\sqrt{S_{NN}}\approx$2-5 GeV).
This pronounced decrease with energy is driven by an increase in
transverse flow. Growing space-momentum correlations 
lead to a shrinking apparent source size. Saturation in flow
velocity~\cite{nuxu} 
(which is reached around the highest available beam energy at the AGS)
leads to a different evolution of the HBT
parameters at the SPS energy range ($\sqrt{S_{NN}}\approx$10 - 20
GeV). One can 
see that transverse HBT radii change very little at SPS. 
The observed behaviour strongly suggests that the HBT correlation
functions directly reflect properties of the collective transverse flow
generated in the collision.\\
%----------------------------------------------------------------------
A consistent picture seems to emerged from the energy dependence systematics of the
pion phase space densities reported during the Conference. 
Preliminary results shown~\cite{laue} by 
the STAR Collaboration seem to confirm the hypothesis of a universal
freeze-out phase space  density~\cite{ferenc} at RHIC energies.
The $P_T$ dependence of the phase space density at midrapidity at RHIC
clearly rules out a static thermal source and is consistent with
a Bose-Einstein distribution  modified to take into account radial
flow. The flow velocity extracted from the analysis of the $P_T$
dependence is quite high ($\approx 0.58c$) with a freeze-out temperature of
about 94 MeV.\\
Study of the freeze-out phase space density by the NA45
Collaboration~\cite{appelshauser} also confirmed a universal phase space
density at 40A GeV. They also reported that the phase space density seems
to be independent of centrality.\\
The E895 Collaboration showed~\cite{lisa} that at AGS energies 
(2-8A GeV) the phase space density distribution is lower than the
``universal'' one and gradually approaches it as energy increases.\\ 
%----------------------------------------------------------------------
It was evident during the Conference that studies of the correlations
of neutral ($\Lambda$, $K^0$) strange particles attracted a lot of
attention. 
Even though the details of the final state interaction of neutral
stange particles are not very well known, which complicates
interpretation of the measurements, the physics which can be accessed
via this type of correlations is clearly attractive.  
In combination with measurements of strange particles yields and
spectra they can provide valuable new information about strangeness
dynamics in the collision.\\
From the experimental point of view such correlations have
certain advantages. Being neutral they pose no Coulomb repulsion
problems, and two-track resolution problems are largely avoided as well.
They are also less influenced by the resonance decays. 
On the other hand such studies pose significant 
experimental challenges since they usually require sizable datasets
and large acceptance detectors due to the relatively low yields and
reconstruction efficiencies of neutral strange particles. Such data were
generally unavailable until now in heavy ion collisions.\\
The NA49 Collaboration presented~\cite{blume} preliminary
analysis of the $\Lambda$-$\Lambda$ correlations from Pb+Pb collisions at
158A GeV. The shape of the measured correlation function shows
a pronounced dip at low relative momentum which is consistent with
Fermi-Dirac statistics of $\Lambda$ hyperons. Also there seems to be
evidence of weak repulsive $\Lambda$-$\Lambda$ potential which may
have some consequences for H-dibaryon searches. This interesting topic
certainly requires further investigation.\\
The STAR Collaboration showed preliminary results of the analysis
of $K^0$-$K^0$ correlations measured in Au+Au collisions at
$\sqrt{S_{NN}}$=130 GeV. The extracted Gaussian radius $R_{inv}\approx6
fm$ is consistent with pion HBT measurements.\\
The E895 Collaboration studied another exotic system - $\Lambda$-p - in
Au+Au reactions at lower AGS energies. Preliminary results~\cite{lisa} are
consistent with an attractive strong interaction between proton and
$\Lambda$ hyperon but cannot be described by the simple source modes
with an Urbana type potential~\cite{wang_pratt}.\\
%--------------------------------------------------------------------
The E895 Collaboration also showed~\cite{lisa} results of the HBT
analysis in the impact-parameter-fixed coordinate
frame. In this approach event-by-event information about
orientation of the reaction plane is combined with information from two-particle
correlations. Such analysis is not only sensitive to the length of
homogeneity and time, but also to the tilt of the source in the
reaction plane.
Experimental access to this level of geometrical detail of the
freeze-out distribution is unprecedented, and represents a new
opportunity to study dynamics of heavy-ion collisions.\\ 
%--------------------------------------------------------------------
\section{Fluctuations}
\label{fluct}
In recent years analyses of the event-by-event fluctuations assumed one
of the central positions in the experimental studies of the
relativistic heavy ion collisions. 
That was evident during the Conference.\\
%----------------------------------------------------------------------
The NA45 Collaboration showed~\cite{appelshauser} analysis of the mean transverse
momentum fluctuations in Pb+Au central collisions at 40, 80 and 158A GeV. They
reported observation of small ($<3\%$), but statistically significant dynamical
fluctuations at all three energies. The observed $\Phi_{P_{T}}$
value~\cite{marek_stashek} is $\approx$8 MeV at 158A GeV. This interesting result
contradicts the small $\Phi_{P_{T}}\approx$ 0.6 MeV/c value reported
earlier~\cite{NA49_fluct} by the NA49 Collaboration.   
Clearly, this discrepancy requires further study.\\
%----------------------------------------------------------------------
The NA49 Collaboration reported~\cite{blume} preliminary results of the
analysis of the event-by-event charge fluctuations in central Pb+Pb
collisions at 40 and 158A GeV. 
They studied fluctuations of ratios of the number of positive and negative
hadrons as a function of the width of the 
pseudorapidity window and applied acceptance corrections following
the procedure suggested by Bleicher, Jeon and Koch~\cite{bjk,koch}. Fully corrected
results are somewhat surprising: fluctuations of charge particle
ratios at both energies are similar and consistent with fluctuations
produced in a pion gas. Naively one would expect at SPS energies
somewhat smaller fluctuations consistent with a resonance gas or perhaps
even with QGP.\\ 
%----------------------------------------------------------------------
The WA98 Collaboration presented results of the search for localized
fluctuations in the multiplicity of charged particles and photons produced
in Pb+Pb reactions at 158A GeV. They reported individual fluctuations in photons
and charged particle multiplicities but no correlated event-by-event
fluctuations of $N_{\gamma}-N_{ch}$. \\
%----------------------------------------------------------------------
The NA44 Collaboration reported~\cite{kopytine} preliminary results of
their search for local fluctuations in the number of charged particles
using a novel event texture analysis. No evidence of critical
fluctuations was found.\\ 
%----------------------------------------------------------------------
For RHIC, the STAR Collaboration reported~\cite{reid} preliminary results of
their extensive event-by-event fluctuations analysis. They presented
an investigation of the charge-dependent and charge-independent $<P_T>$
fluctuations individually, as well as the overall $<P_T>$ fluctuations.
A significant excess in fluctuations of $\Delta\sigma_{P_T}\approx35$ MeV/c
beyond the statistical expectation was observed in the
charge-independent analysis. Also, a somewhat smaller but still
statistically significant excess was observed in the charge-dependent
analysis. 
They also studied the centrality dependence of these excess
fluctuations where hints of a decrease with centrality in
the charge-dependent part was observed.
The results of multiplicity fluctuations analysis show no evidence for
predicted fluctuation suppression associated with rapid hadronization
of the QGP. \\
%----------------------------------------------------------------------
Event-by-event fluctuations also drew a lot of attention
from theorists. Most of the theoretical studies reported at this
Conference were devoted to the fluctuations of so-called ``conserved
quantities'': charge, baryon number, strangeness, etc.\\ 
Koch discussed the theoretical situation with conserved quantities fluctuations in
general and charge ratios in particular. He pointed out effects of
limited detector acceptance and the importance of related corrections.
It was apparent
during the Conference that current theoretical interest is aimed at the
studies of the ``survivability'' of the proposed observables in the
real world situation. Most of the results were devoted to the studies
of the influence of the system expansion and limited experimental acceptance.
These were discussed in talks by Asakawa~\cite{asakawa},
Stephanov~\cite{stephanov}, Gavin~\cite{gavin}.\\
Pratt~\cite{pratt_bf} presented a new observable for determination
of the nature of hadronization in heavy ion collision - the balance
functions. The balance function describes the conditional
probability that a particle in the momentum bin $p_1$ will be
accompanied by a particle of opposite charge in the bin
$p_2$~\cite{bass_danielewicz_pratt}. Balance 
functions probe hadronization by quantifying charge correlations in
momentum space and potentially allow us to distinguish scenarios with
QGP formation from purely hadronic ones. First experimental attempts
to apply this observable to heavy ion collisions are underway. 

\section{Conclusions}
\label{conc}
Many new and interesting HBT results have appeared since last Quark
Matter. Combined results from the AGS, SPS and RHIC allow us to have a
systematical view of the evolution of the HBT parameters with beam energy.
Presentations of the first results from RHIC experiments were some of
the highlights of the conference. The most intriguing feature of the
preliminary HBT results from RHIC is $K_T$ dependence of the $R_O/R_S$
ratio reported by the STAR Collaboration. It is not clear yet what
physics can lead to such a dependence.\\
It would be interesting to perform measurements between AGS and SPS
energy domains, which seems to be a transition region in terms of HBT
systematics.  
It would be also useful to probe the energy range between SPS and RHIC. 
Such measurements will significantly extend our understanding of
spatio-temporal evolution of heavy ion collisions.\\
The situation with fluctuation measurements at SPS is far from clear and further
experimental and theoretical efforts are needed to resolve existing ambiguities. \\
Preliminary fluctuations results from RHIC are tantalizing and
warrant further investigation.

\end{document}